%% file: main.tex
\documentclass[pra,showpacs,aps,amssymb,amsmath,nofootinbib,notitlepage,superscriptaddress,twocolumn,longbibliography]{revtex4-2}

\usepackage{graphicx,graphics,epsfig,times,bm,mathrsfs}
\usepackage[normalem]{ulem}
\usepackage{subfigure}
\usepackage{xfrac}
\usepackage{mathtools}
\usepackage{amsmath,amssymb,amsfonts}
\usepackage[table, xcdraw]{xcolor}
\usepackage{varwidth}
\usepackage{babel}
\usepackage{hhline}
\usepackage{multirow}
\usepackage{longtable}
\usepackage{pgf}
\usepackage{physics}
\usepackage{bbm}
\usepackage{pgfplots}
\pgfplotsset{compat=1.17} 
\usepackage{listings}
\usepackage{paralist}
\usepackage[hidelinks]{hyperref}
\usepackage{float}
\usepackage{microtype}
\usepackage[T1]{fontenc}
\usepackage[utf8]{inputenc}
\usepackage{algorithm}
\usepackage[noend]{algpseudocode}

\pgfdeclarelayer{bg}    
\pgfsetlayers{bg, main}  

\algnewcommand\algorithmicforeach{\textbf{for each}}
\algdef{S}[FOR]{ForEach}[1]{\algorithmicforeach\ #1\ \algorithmicdo}
\algrenewcommand\algorithmicindent{0.55em}%

\newcommand{\enquote}[1]{``#1''}

\makeatletter
\newcounter{protocol}

\makeatother

\begin{document}

\title{Packet Switching in Quantum Networks: A Path to Quantum Internet}

\author{Stephen DiAdamo}
\email{sdiadamo@cisco.com} 
\affiliation{Cisco Research, Garching bei M\"unchen, Germany}

\author{Bing Qi}
\email{bingq@cisco.com}
\affiliation{Cisco Research, Los Angeles, CA, USA}

\author{Glen Miller}
\affiliation{Acacia Communications, Mass-Scale Infrastructure Group, Cisco, Maynard, MA, USA}

\author{Ramana Kompella}
\affiliation{Cisco Research, San Jose, CA, USA}

\author{Alireza Shabani}
\email{ashabini@cisco.com}
\affiliation{Cisco Research, Los Angeles, CA, USA}

\date{May 16, 2022}

\begin{abstract}

Large-scale quantum networks with thousands of nodes require scalable network protocols and physical hardware to realize. In this work, we introduce packet switching as a new paradigm for quantum data transmission in both future and near-term quantum networks. We propose a classical-quantum data frame structure and explore methods of frame generation and processing. Further, we present conceptual designs for a quantum reconfigurable optical add-drop multiplexer to realize the proposed transmission scheme. Packet switching allows for a universal design for a next generation Internet where classical and quantum data share the same network protocols and infrastructure. In this new quantum networking paradigm, entanglement distribution, as with quantum key distribution, is an application built on top of the quantum network rather than as a network designed especially for those purposes. For analysis of the network model, we simulate the feasibility of quantum packet switching for some preliminary models of quantum key and entanglement distribution. Finally, we discuss how our model can be integrated with other network models toward a realization of a quantum Internet.

\end{abstract}

\maketitle

\section{Introduction}\label{sec:intro}

Quantum computing, networking (communication), and sensing are emerging as frontier technologies for information processing. Quantum computers can offer better solutions to hard problems in chemistry and material science \cite{Cao2019,Bauer2020}, as well as in machine learning \cite{Liu2021,Huang2021}. Quantum networks enable quantum-secure communication \cite{xu2020secure,Pirandola2020} and entanglement-assisted communication \cite{bennett1992communication}, while higher measurement sensitivity can be achieved with quantum sensors \cite{Degen2017}. Among quantum information technologies, quantum networking in particular is finding a more central role. Beyond comm\-unica\-tion and cryptography related applications, quantum net\-working can be a key part of developing large-scale quantum computers via interconnecting multiple quantum chips \cite{Monroe2014, IBM2020roadmap, DiAdamo2021}. Also, networked quantum sensors exchanging quantum information collectively have stronger sensing power \cite{Zhang2021,Bringewatt2021}.

One can envision for the future thousands, or even millions, of devices connected over a global \textit{quantum network of networks}---namely, a quantum Internet. Which networking infrastructure and protocols can allow for the realization of such a large-scale quantum network? Despite previous proposals \cite{illiano2022quantum} there has been no comprehensive answer to this question and proposals for quantum networking protocols and network layers have been limited to particular applications such as entanglement distribution \cite{wehner2018quantum, pirker2019quantum, alshowkan2021lessons} or quantum key distribution (QKD) \cite{alia2021dynamic}. Also from the infrastructure perspective, there has been a quest for classical and quantum network coexistence \cite{townsend1997simultaneous, alia2022dv, geng2021coexistence}, although with no clear coexisting protocols or network specifications. In this paper, we introduce a packet-switched quantum network as an answer to the above question. Before presenting our proposal for packet switching it is important to discuss the bigger picture and how we conclude packet switching as the path forward. We start by introducing the following basic engineering principals which need consideration when designing a large-scale quantum network, and specifically, a quantum Internet: 

1) \emph{Universality:} We define a universal quantum network as one which can accommodate any application, and not just those which rely on entanglement distribution \cite{wehner2018quantum, pirker2019quantum, alshowkan2021lessons}. A prominent example is the BB84 prepare-and-measure QKD protocol \cite{Bennet84} which does not require any entanglement resources. In this regard, a future quantum Internet should serve all users universally, independent of the use case.  Correspondingly, quantum networking protocols and layers should serve any application whether it is QKD, entanglement distribution, blind computing \cite{Fitzsimons17}, or any other application. Note that here we consider entanglement distribution (and correspondingly teleportation) as an application rather than a layer of quantum networks.

2) \emph{Transparency:} Quantum networks can use the same physical media as classical optical networks for data transmission, whether optical fiber or free space links. Although quantum networking experiments often employ separate fibers for classical and quantum signals, we consider any realistic field deployment to rely on the coexistence of classical and quantum signal over the same medium. Furthermore, we assume as an engineering principle that future Internet protocols should function regardless of the nature of transferred data, whether it is classical or quantum. Therefore, we believe quantum-classical coexistence is not merely sharing the same physical media \cite{townsend1997simultaneous} but sharing the same network routing protocols, layers, and standards.

3) \emph{Scalability}: The quantum networking protocols should be able to support the growth of the network. In the classical world, the primary reason for the development of packet switching in telecommunication systems was to accommodate the growing number of users and need for high-bandwidth communication. Packet switching allows easier network management, dynamic correction for the points of failure, and better use of physical infrastructure. Moreover, much of the success of classical communication networks can be accredited to the standardization approaches used for scaling and interoperability. With standardization, networks composed of software and hardware developed by various vendors can be integrated into existing networks with performance guarantees at very little effort. We therefore believe with an early adoption of packet switching for quantum networks, we prevent a later transition from a collection of switching approaches seen today to one which can allow for easier standardization, using the same already-carved path classical networks used.

We consider universally, transparency, and scalability as three design principals for a quantum Internet, or simply the design principles for the future Internet. In this work, we introduce packet switching as a quantum networking paradigm that satisfies these principles. 

This article is organized as follows: in Section~\ref{sec:switching}, we introduce our approach to packet switching in an optical quantum network, also introducing our classical-quantum hybrid frame structure and a proposal for optical network hardware to implement the approach in Section \ref{sec:hardware}. In Section \ref{sec:applications}, we investigate two example use-cases for our network model, namely quantum key and entanglement distribution, appropriate for a near-term implementation of our network model. In Section \ref{sec:layers}, we investigate how our network model can be integrated with currently deployed optical networks and moreover with currently deployed or proposed quantum networks. The paper is concluded in Section \ref{sec:conclusion} along with final discussion and outlook.

\section{Packet Switching for Optical Quantum Networks}\label{sec:switching}

The first communication networks installed commercially relied on the concept of circuit switching in order to route data packets in the network \cite{roberts1978evolution}. At a high level, circuit switching is a switching approach that, given a route in a network, reserves the entire capacity of the channels in the route until communication is terminated. Indeed in some situations, circuit switching can provide advantages, such as maintaining synchronization, low latency transmissions, and overall simplicity. One major downside of circuit switching was that in scenarios of bursty network traffic, the communication channels could often be reserved with little to no traffic being transmitted. To improve the utilization of the network, and to aid in accommodating the growing number of network users, the concept of packet switching was widely adopted~\cite{roberts1978evolution}.

Given that the concept of quantum networks is in a very early stage of development, there has not yet been much consideration on how to handle many users in the network. In particular, of the current proposals for quantum networks, there has not been an equivalent design proposed analogous to packet switching. Indeed currently deployed quantum networks need only to support very few users and can only supply low communication rates. The largest reported entanglement-based QKD network for distributing quantum keys without trusted nodes was demonstrated in \cite{joshi2021protocols}, supporting 8 unique users with a average of 1 kb/s of secure key in the best case. The approach used in \cite{joshi2021protocols} was not a routing approach with packet switching in a networking sense, but rather an assignment of unique wavelengths assigned for each user allowing for communication between any two users, essentially dedicating a direct communication line between pairs of users. In this sense, the 8-user network forms a complete network such that no link-layer functionality for switching is involved. In \cite{chen2021implementation}, a 46-node QKD network is demonstrated with switches and trusted nodes at 49.5 kb/s, and in the work, a circuit-switched approach is used. 

For future quantum networks, we believe packet switching will allow for many users as well as support higher communication rates, but the construction of such a system will naturally come with its own challenges. The implementation of packet switching in quantum networks can be largely inspired by classical optical networks, but indeed there are major challenges to overcome due to the nature of quantum systems in comparison to purely classical systems. In optical networks, it is common that when a packet arrives, the control information and payload of the packet is converted from the optical domain into the electrical domain. This implies the optical signal is measured and converted into a computer-readable format. When the packet is ready for retransmission, the payload is converted back into the optical domain and sent onward through the next fiber. This process is known as O/E/O conversion. During this waiting time, the payload information can be stored essentially for as long as necessary without any deterioration to the signal, and moreover the information can be corrected for any errors. As network traffic rates increased, and because the operation of converting optical signals to electrical signals was---and in some cases still is---expensive to perform, or the computational power was missing, methods for keeping the payload information in the optical domain were introduced \cite[Chapter 13]{keiser2021fiber}. Devices such as optical switches, wavelength selective switches, reconfigurable add drop multiplexers, and other optical components were invented \cite{saleh2012all}. The problem with these devices is that they further attenuate the signal. With classical information, the payload signal can be amplified to mitigate signal attenuation due to the components, maintaining a strong signal for the next transmission in the route \cite{saleh2012all}.

In the quantum case, many of these solutions do not carry over. The nature of quantum mechanics is such that arbitrary quantum states cannot be measured to then be regenerated in the same state. Quantum states can be in a superposition of states and measuring the system forces it to collapse into one basis state, destroying the superposition and thereby any encoded quantum information. Without the full knowledge of the quantum state before measurement, the state cannot be perfectly regenerated, and so conversion of a quantum state to the classical domain is not possible. This prevents us from using classical data storage and error correction approaches. Indeed with optical memories \cite{lvovsky2009optical, arnold20211}, a similar approach could be considered. Another problem to overcome is that quantum states are very fragile, and storing them in memories can rapidly cause the decoherence of the quantum state, ruining the encoded information. Moreover, the writing and reading steps for storing and retrieving a quantum state from memory can cause additional loss on an already inherently weak quantum signal. With these known effects, the amount of time that a quantum state remains in a memory should be minimized. 

Performing error correction on a quantum state while it is in memory could also be possible, but would require a quantum computer capable of performing the necessary operations, a technology that is expected in the future. This level of technology is essentially an all-optical quantum repeater~\cite{azuma2015all}. Using the hardware components for all-optical classical networks is the more near-term approach to overcoming issues with storing quantum systems, but another hurdle exists that is non-trivial to overcome. Many existing optical network technologies rely on signal amplification to overcome attenuation effects. Because arbitrary quantum states cannot be copied \cite{wootters1982single, dieks1982communication}, amplifying weak quantum signals is not a possible method for overcoming attenuation. Therefore, one would need to construct low-loss optical hardware for improving the performance and reliability of the network.

Optical communication is the backbone of telecommunication systems, where data is encoded into laser pulses and transferred over optical fibers or free space. In its current form, optical communication is not purely optical as it transforms back and forth into the electrical domain at intermediate nodes for error correction and amplification \cite[Chapter 13]{keiser2021fiber}. In early days of optical communication, building an all-optical network where data stays in optical domain all along the network was a goal \cite{saleh2012all}. Realizing an all-optical network remains an unachievable goal due to challenges not much different from what we face in quantum networks, where the main difference is that instead of the strong light signals used in classical communications, we have weak light signals that cannot be copied. Although a rather obsolete field, we can still learn from the older works in this area.

Overall, the path to building large-scale quantum networks capable of supporting a large number of users will require both the invention of new technology as well as optimizing current technology to accommodate the challenges of transporting quantum systems. With these challenges in mind, we present an approach to packet-switched quantum networks that is both future-proof, accounting for the arrival of novel quantum technologies, as well as timely, in the sense that in the near-term our approach can already be used to deploy some quantum networking applications. In this section, we aim to learn from history and thus define a packet switching approach preemptively to prepare for the coming of large-scale quantum networks, supporting the deployment of many quantum network applications as well as many users.

\subsection{Classical-Quantum Hybrid Frame Structure}

\begin{figure}
    \centering
    \includegraphics[]{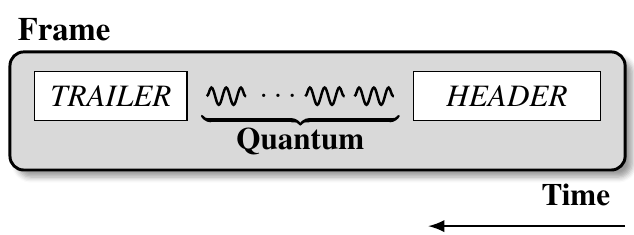}
    \caption{Quantum frame structure. The frame is composed of three components: A classical header, a quantum payload, and a classical trailer.}
    \label{fig:frame-structure}
\end{figure}

\begin{figure}
    \includegraphics[]{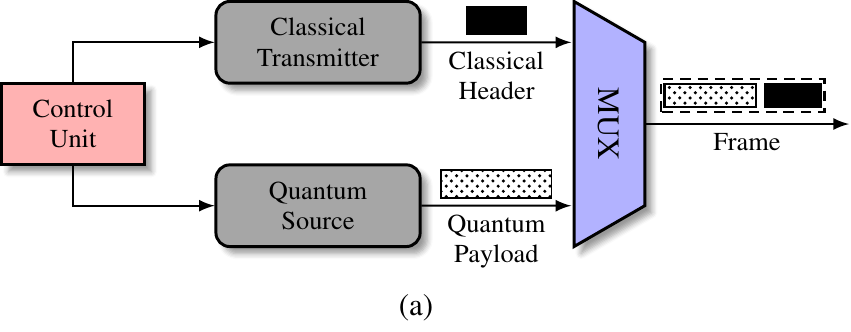}
    \includegraphics[]{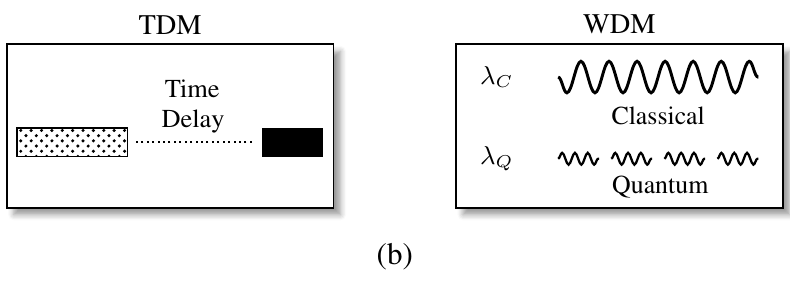}
  \caption{Hybrid frame generation. (a) The classical header and quantum payload can be generated from different photon sources and multiplexed into a hybrid data frame using a MUX. (b) Two examples of multiplexing schemes: time-division multiplexing (TDM) and wavelength-division multiplexing (WDM).} 
  \label{fig:TDMWDM_new}
\end{figure}

In order to perform packet-switching in a quantum network, we need to firstly define a data frame structure capable of handling a quantum payload. The central idea we use is a hybrid frame structure: each quantum payload is framed with a classical header and trailer. The header contains crucial information for routing, error mitigation and correction. The trailer indicates the end of the quantum signal as seen in Fig.~\ref{fig:frame-structure}. The classical header and trailer and the quantum payload can be generated using different photonic sources and multiplexed into a hybrid data frame by using a different degree of freedom of light, such as time, wavelength, polarization, spatial mode, and so on, or any combinations of them, as shown in Fig.~\ref{fig:TDMWDM_new}. 

In Fig.~\ref{fig:TDMWDM_new}~(a) is a depiction of a source node which generates a hybrid frame. A control unit is used to trigger hardware components that generate signals for the classical header and trailer as well as the quantum payload. The classical transmitter encodes the header and trailer data and a quantum source emits and encodes quantum states. The two signals are then multiplexed using some predetermined form of multiplexing and sent through the fiber. The quantum source depicted is a general source and can emit arbitrary quantum states, even multi-mode systems. In the figure, we do not depict the trailer portion of the frame, but the trailer will follow the quantum payload as depicted in Fig.~\ref{fig:frame-structure} using the same encoding and multiplexing as the header. Fig.~\ref{fig:TDMWDM_new}~(b) is a depiction of two multiplexing approaches. On the left is time division multiplexing, where the header information will be sent earlier in time than the quantum payload, so that the receiver can distinguish the two. On the right is wavelength division multiplexing, where the signals can be sent simultaneously but at different wavelengths.

In our proposal for a quantum network data frame, we modify the contents of the header and trailer to attempt to accommodate the uniquely-quantum noise effects that arise. In this preliminary work, it is difficult to plan for the coming novel technologies surrounding quantum networks. As a first step, we modify the standard Link-Layer Discovery Protocol frame (LLDP) \cite{lan2009ieee} to accommodate quantum data transmission, forming a quantum Ethernet frame. The motivation behind this is that LLDP frames are widely used and integrated with existing switching hardware. Moreover, LLDP frames have flexibility to add additional fields for switching choices and error correction. The general contents of our frame header are much like a traditional frame, it should contain information regarding the source and destination and specific control codes. On top of the parts found in standard LLDP frames, we add additional fields to the Type-Length-Value (TLV) part of the frame that are specifically quantum. In this stage of development, fields we imagine will be the details surrounding quantum error correction and mitigation. Data such as how long the quantum payload has spent in a quantum memory, a maximum cut-off time until the frame is dropped, and the quantum error correction protocol to be used for error correction should be integrated. For the trailer, we use the same frame structure with a binary value in a Header/Trailer TLV to distinguish the two.  We detail the frame in Table~\ref{table:lldp}. At this stage, we imagine the trailer will simply indicate the end of the signal, but future iterations could integrate trailer structures that incorporate additional resources for quantum channel estimation \cite{ji2008parameter}, or other error correction information for example. Future work will be to form concrete structures for the header and trailer to work in a completely general setting.

\setlength{\tabcolsep}{0.5em} 
\renewcommand{\arraystretch}{1.2}

\begin{table*}
\centering
\includegraphics[]{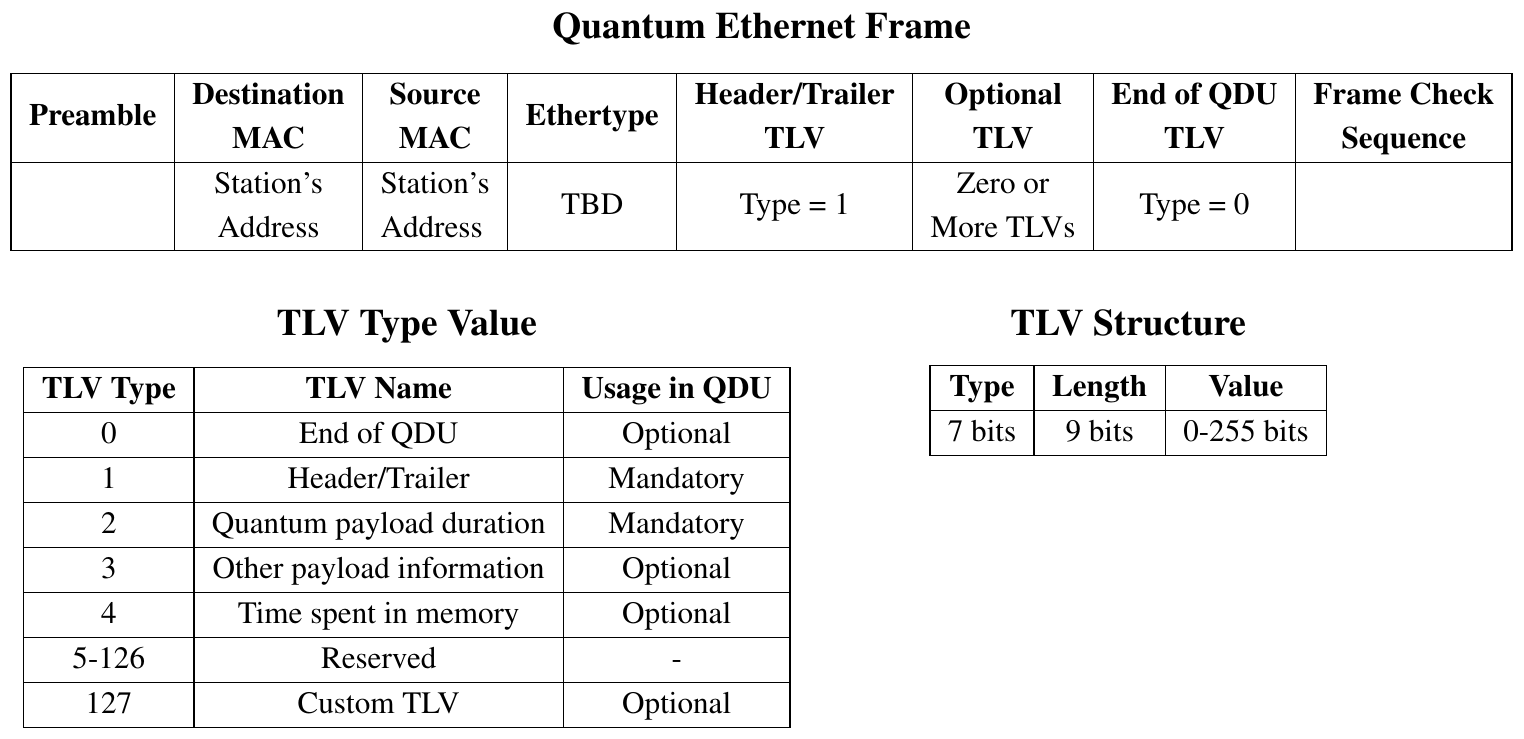}
\caption{The modified LLDP frame structure for quantum data transmission. Here we add additional Type-Length-Value (TLV) data for quantum and consider a Quantum Data Unit (QDU).}
\label{table:lldp}
\end{table*}

\subsection{The Approach to Packet Switching}

With a frame structure defined, the question of how a frame is transmitted and processed at relay nodes is next to be considered. We have discussed how the classical and quantum information can be multiplexed, but how a frame is logically processed and the type of hardware available at each relay switch can dictate the precise timing of the frame generation. In a classical setting, because it is possible to amplify, copy, and correct errors, the precision of timing for when the parts of a frame are transmitted is less critical---a frame payload can be sent with its header shortly in front of it with a closely following trailer. With quantum information, because of the complexity involved in transporting and storing it, precise timing and synchronization are critical. For this, we propose the use of two switching schemes based on classical schemes used in optical networks. 

\begin{figure}
    \centering
    \includegraphics[]{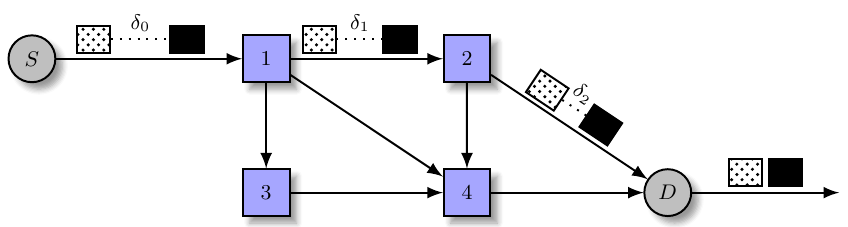}
    \caption{An example of burst switching. A frame is sent from node $S$ to node $D$ in the network with the header part of the frame (solid) sent $\delta_0$ seconds earlier than the payload. Once the frame passes through node $1$, the time-gap between the header and the payload decreased to a time difference of $\delta_1 < \delta_0$, due to the processing delay at node $1$. This repeats at node $2$ and at the destination $D$, where the frame is dropped from the network.}
    \label{fig:burst-switch}
\end{figure}

Technology for quantum memories is in a very early stage of development and the capability of the memories for maintaining quantum state fidelity is generally in the millisecond to low single digit seconds range, especially lower in systems coupled to networks \cite{cho2016highly, abobeih2018one}. We therefore propose a near-term approach that makes no use of quantum memory, appropriate for an initial implementation stage of our quantum network model. Our scheme is based on a switching scheme known as burst switching, or \enquote{just-in-time} switching. The idea here is, given roughly the number of hops from source to destination, to add enough guard time---the time between when the header is transmitted and when the quantum payload is transmitted---such that at each node, a routing decision can be made before the payload arrives. Once the payload arrives, if processing is complete, then the payload can be sent onward. If processing is not complete, or the node is not able to handle the frame, the payload is discarded and a retransmission can be made with more guard time. Depicted in Fig.~\ref{fig:burst-switch} is an example of burst switching. In a sense, this approach is a hybrid approach between packet switching and circuit switching. The benefit gained in this case over circuit switching is that dynamic routing is still possible. This additional freedom can therefore allow for better network utilization.

In the next case, we assume the network nodes are capable of storing quantum payloads. Depicted in Fig.~\ref{fig:q-switch} shows an initial hardware concept. In this case, we do not require as precise a guard time, and the stronger the memory, the weaker the precision needs to be. When the header arrives at a node, if by the time the payload arrives the node is busy, the payload will be switched into a quantum memory to be stored until transmission becomes possible. Here, for example, a node can make use of the optional parameters of the proposed quantum Ethernet frame that can state a maximum storage time for the quantum payload before it should be dropped. If the payload requires a higher fidelity and the maximum duration for storage has elapsed, the payload can be dropped. 

In both cases, while the classical header and trailer can be amplified, measured, and regenerated at each network router, the quantum payload should be transmitted through the entire network faithfully without being measured or amplified at the intermediate network nodes. Our proposals are conceptually similar to the photonic packet switching in classical all-optical networks \cite{yao2000advances, yoo2006optical}, so there are some common challenges in both quantum and classical schemes. We will explore ways to overcome these challenges in detail in future work.

\section{Network Hardware Requirements for Packet Switching}\label{sec:hardware}

With the frame processing defined in the previous section, we explore concepts for the physical hardware capable of processing the hybrid frames. One key building block of packet-switched quantum networks is a quantum version of Reconfigurable Optical Add Drop Multiplexer (ROADM)---a q-ROADM---which acts as a quantum switch able to route quantum information through a large network. We remark that a QKD network using a modified classical ROADM has been demonstrated recently \cite{wang2019end}, where the authors bypassed the optical amplification unit and reduced the insertion loss for the QKD signals. While this approach can be useful in near-term applications like QKD, it is not adequate for a general-purpose quantum network due to the lack of crucial quantum functionalities such as payload storage and dynamic switching. Furthermore, it is designed for circuit-switched networks, rather than packet-switched networks discussed in this paper.

\begin{figure}
    \centering
    \includegraphics[]{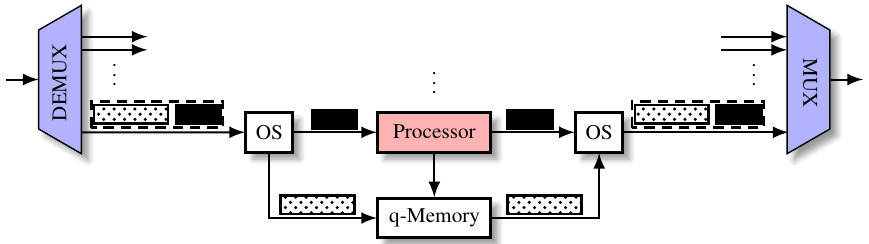}
    \caption{A first iteration device for quantum frame processing and switching. The classical-quantum hybrid frames enter a demultiplexer and the header (solid) is split from the quantum payload (dotted) via an Optical Switch (OS). The header is processed into an electrical domain and the information used to determine the next hop. Once determined, the quantum payload is released from the q-Memory, merged with the header by another OS, multiplexed with the other frames, and sent onward.}
    \label{fig:q-switch}
\end{figure}

In this section, we consider two paths to realizing a hardware capable of processing the classical-quantum hybrid frames. On one path, we aim for a near-term solution, capable of processing a frame with limited additional functionality, and on the other, a longer-term design with additional components that will make the network more robust and efficient. In the first case we target only the essential features required to process the frame, ignoring error correction, signal frequency conversion, or any other optimizations that can complicate the system. We depict the hardware design in Fig.~\ref{fig:q-switch}. In this case, the multiplexed classical-quantum frame enters a de-multiplexer to separate the header information from the payload. The header enters a processor, and the quantum payload a memory. The memory in this case can be a delay line or a more robust optical memory. The classical information is processed to determine when to release the quantum payload and then is regenerated and multiplexed with the payload. The multiplexed frame is sent through an optical switch to be sent to the next destination, along with the classical trailer following, processed the same way as the header. In general, in this setting the transmission protocols should be designed with very limited quantum memory abilities in mind, as explained in the previous section.

\begin{figure*}
    \centering
    \includegraphics[]{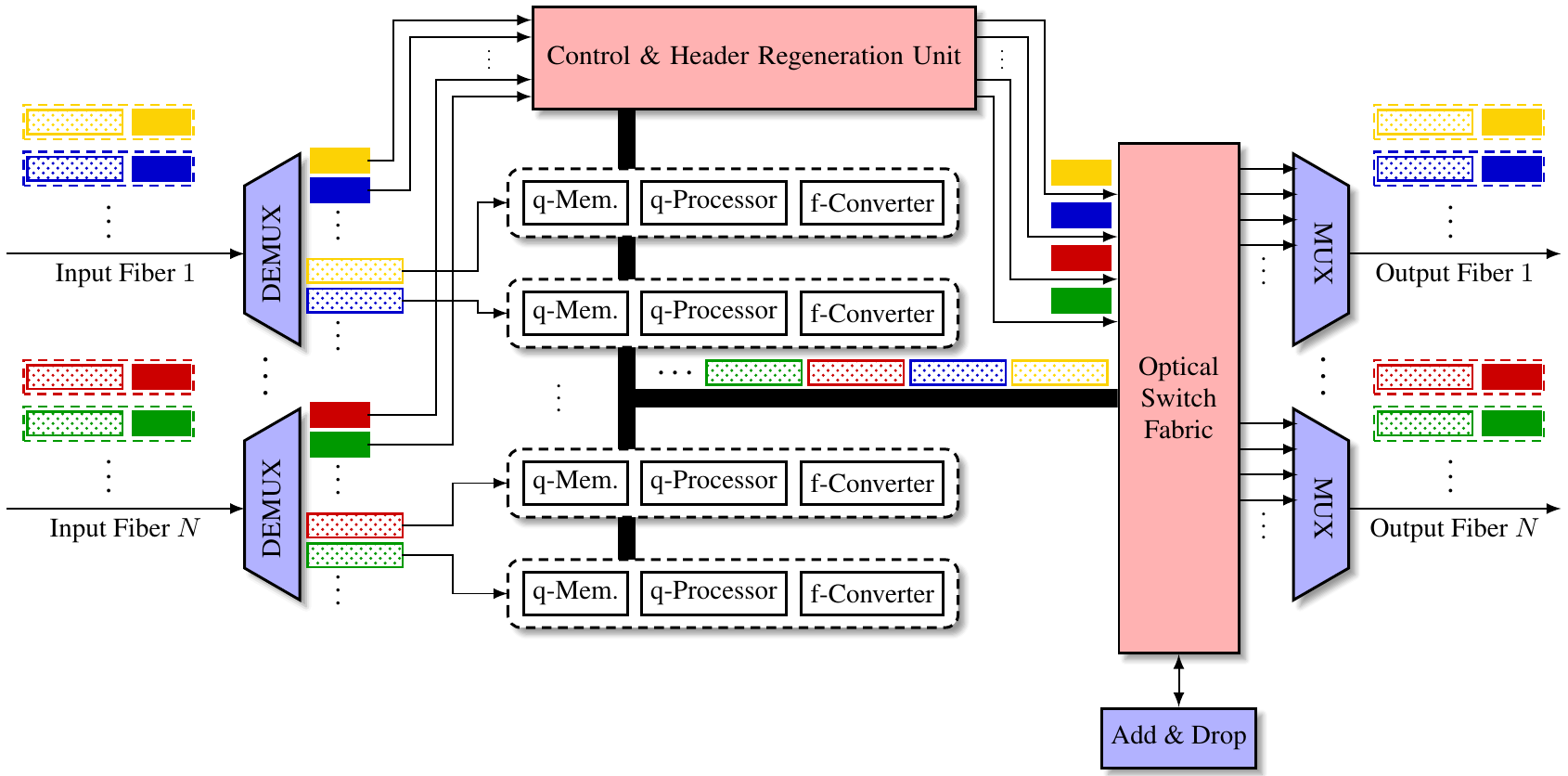}
    \caption{A conceptual design of a quantum ROADM. From the left side, $N$ fibers containing multiplexed frames enter a demultiplexer to separate the classical header information from the quantum payload. The classical header if processed in the Control and the quantum payload sent to a quantum processing unit. In the quantum processing unit, the quantum payload is stored, corrected of errors, and if needed, undergoes frequency conversion. Once ready for transport, the payload travels through optical switch fabric and it output behind its regenerated header.}
    \label{fig:qROADM}
\end{figure*}

For the second, more distant future, path, inspired by the classical ROADM, a conceptual design of a q-ROADM is shown in Fig.~\ref{fig:qROADM}. This design incorporates more complex components that can achieve error correction and signal frequency conversion. It also is designed to handle many users. Here, a q-ROADM can have multiple input fibers, output fibers, and add/drop channels, similar to a classical ROADM. Each fiber may carry multiple wavelength channels. Depending on the multiplexing scheme employed, a corresponding de-multiplexing is used to separate the classical header and the quantum payload. After being optically amplified and converted into electrical signals, the classical information contained in the header will be fed into a control unit, to decide how to further process the quantum payload. In the meantime, the quantum payload will be stored in quantum memory (q-Memory), and the classical trailer will arrive, following the same path as the header. Depending on the application, quantum error correction or other quantum operations may be performed on the quantum payload by the q-Processor. Based on the state of network traffic, the control unit will determine the optimal output, and provide control signals to release the quantum payload from q-Buffer and perform wavelength conversion if needed. It also regenerates the classical parts of the frame at the suitable wavelength and provides the control signal to the optical switch fabric to multiplex the header with the quantum payload and route the hybrid data frame to the suitable output fiber.

A fully functional, general-purpose, q-ROADM for a packet-switched quantum network may require significant breakthroughs for many fundamental quantum technologies, such as long-lived quantum memory and quantum repeaters. In the meantime, we can make use of near-term technologies and trade-off with a focus more on the timing and synchronization in order to make do with presently available hardware. Indeed, with a near-term approach, we cannot overcome the attenuation effects of the optical components. As discussed earlier, quantum networking has its unique challenge associated with the quantum no-cloning theorem \cite{wootters1982single, dieks1982communication}, where an arbitrary unknown quantum state cannot be perfectly cloned. This implies that, with the exception of some loss-tolerant protocols like QKD, quantum repeaters \cite{munro2015inside} capable of distant entanglement generation or error correction are required for constructing general-purpose large-scale quantum networks. 

\section{Example Applications Built on Packet-Switched Quantum Networks }\label{sec:applications}

In the near term, without the ability to store quantum information for long periods nor the ability to perform robust quantum processing, overcoming the effects of loss for long distance quantum communication is not possible for arbitrary quantum states. There are, however, some applications that can still function under heavy loss. In this section, we explore an initial analysis of two such applications; quantum key distribution and entanglement distribution.

\subsection{Quantum Key Distribution}
One of the most mature quantum communications protocols is Quantum Key Distribution (QKD). QKD is a protocol that allows two remote users (e.g. Alice and Bob) to establish a secret key using an untrusted quantum channel and an authenticated classical channel \cite{xu2020secure, Pirandola2020}. The generated secret key can be further applied in various cryptographic protocols to achieve information-theoretic security, using encryption schemes like the one-time-pad. Alternatively, other secure encryption schemes like the Advanced Encryption Standard (AES)~\cite{daemen2001reijndael}, which requires less key material generally consuming 128 or 256 bits, and refreshing periodically, could be employed.

There are two key features that make QKD much easier to implement than other quantum network protocols, which can require transferring arbitrary quantum states, such as distributed quantum computing. First, most QKD protocols are loss-tolerant in the sense that a secret key can still be generated even when most of the transmitted photons are lost in transmission \cite{xu2020secure, Pirandola2020}. This suggests that the storage of QKD signals in a network switch may not be necessary and moreover can be routed through imperfect optical devices such as optical switches. Second, most QKD protocols are noise-tolerate in the sense that secret key material can be generated from quantum signals of relatively low fidelity by performing classical error correction and privacy amplification.

So far, all existing QKD demonstrations have been based on the circuit switching scheme, where a dedicated quantum link is established between the sender and receiver before quantum transmission starts. However, such a scheme may not be efficient in a large-scale quantum network in a similar way as was experienced with early telephone communication networks. As discussed in Section \ref{sec:switching}, a hybrid quantum-classical data frame could be constructed by adding a classical header (trailer) in front of (behind) the quantum payload. On one hand, the classical header is structured to carry information such as the network addresses of the sender and the receiver, the data structure of the quantum payload, etc., and can be measured, processed, and regenerated at intermediate networks nodes. On the other hand, to maximize the QKD performance, disturbance to the quantum payload should be minimized along the quantum channel and at the network nodes.

We remark that the security of QKD is not dependent on the details of the quantum channel or routing algorithms. In fact, in standard QKD security proofs, the quantum channel is assumed to be fully controlled by the adversity (Eve). This is convenient because we do not need to develop a new security proof for packet-switched QKD networks. Nevertheless, the performance of QKD, in terms of secret key rate and QKD distance, is highly dependent on both the quantum channel characteristics and routing strategies. It is therefore important to conduct numerical simulations to study the performance of QKD in different network configurations.

Among various QKD protocols, in this paper, we study a representative prepare-and-measure protocol, namely the BB84 protocol \cite{Bennet84}. The central idea can also be applied to other QKD protocols, such as entanglement-based QKD \cite{ekert1992quantum}, measurement-device-independent QKD \cite{lo2012measurement}, and others. More specifically, we consider the polarization encoding BB84 QKD, where Alice prepares single-photon states with polarization randomly chosen from $\{H, V, D, A\}$, where $H$ ($V$) refers to horizontal (vertical) polarization state and represents bit $0$ ($1$) in the rectilinear ($Z$) basis, while $D$ ($A$) represents $45$ ($135$) degree polarization state and represents bit $0$ ($1$) in the diagonal ($X$) basis.

For transmission, Alice, the sender, labels $k$ encoded single-photon pulses into a frame by adding a classical header, emitting the $k$ photons, adding the trailer, and proceeds to send the hybrid data frame to Bob, the receiver, through a packet-switched quantum network. Alice repeats the above process, until all $m$ desired frames are transmitted. At Bob's end, he processes the classical header and trailer and the quantum payload separately. The classical parts of the frame can be measured with a classical detector and provide Bob with the information about the sender and the quantum payload. The polarization of each photon in the quantum payload is then measured in a randomly chosen basis. After the $m$ frames have been registered, or a maximum QKD session time has been reached, Bob informs Alice and they start the post-processing protocol using a classical authenticated channel, which could be over a separate classical network, or in our vision of quantum networks, over the same channel.

In the classical post-processing stage, Alice and Bob use data collected in $X$ basis to upper bound Eve's information and data collected in the $Z$ basis for secret key generation. For simplicity, we further assume that Alice has a perfect single photon source. The asymptotic secret key rate (per transmission) for the efficient BB84 QKD protocol \cite{lo2005efficient} is given by, 
\begin{equation}\label{eqn:secret-key-rate}
    R=KQ[1-f \cdot H_2(e_Z)-H_2(e_X)],
\end{equation}
where $Q \in [0, 1]$ is the gain, $e_Z$ and $e_X$ are Quantum Bit Error Rate (QBER) in $Z$ and $X$ basis correspondingly, and $H_2(x)=-x\log_2(x)-(1-x)\log_2(1-x)$ is the binary Shannon entropy function. The parameter $f$ quantifies the inefficiency of the classical key reconciliation process. Here, we also introduce a parameter $K \in [0, 1]$, which quantifies the data loss due to the routing strategy, as will be discussed below.

\begin{figure}
    \includegraphics[]{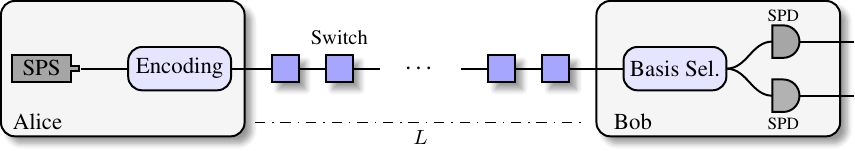}
  \caption{A simple network configuration for QKD simulation. The sender has a Single Photon Source (SPS) which leads into an encoding device which prepends header information and appends a trailer information. The frame is encoded according the BB84 protocol in this scenario. The encoded frame is then sent into the network where it will travel through a number of switches. At each switch there is a probability of data drop due to network contention and some loss due to the header processing time. Eventually, the frame arrives at the receiver where  the standard BB84 decoding scheme is performed using a basis selection  device which directs photons to one of the two Single Photon Detector (SPD).}
  \label{fig:QKD_model}
\end{figure}

In a real QKD process, all the parameters in Eq. \eqref{eqn:secret-key-rate} can be estimated from experimental results without any knowledge about the quantum channel. However, to estimate the QKD performance by numerical simulation, we establish a theoretical model to describe both the QKD system and the quantum channel. The model depends on both the network configuration and the routing algorithm. As an illustrative example, we consider a simple network configuration, where the two QKD users are connected by a single path with $n$ intermediate switches, as shown in Fig.~\ref{fig:QKD_model}. In this analysis, we assume the switches are early-stage quantum switches and have no quantum memories, and therefore no storage ability. Each switch can separate the classical parts of the frame and quantum payload, read out the classical information, regenerate the classical header and trailer, and detour the quantum payload to the outgoing channel if available. To introduce some network dynamics, we assume the outgoing channel at each switch is available with a probability of $P$ when the QKD signal arrives.

We adopt a simple routing strategy as follows: (1) If the channel is available, the quantum payload goes through with a fraction of loss corresponding to the processing time of classical header; (2) If the channel is not available, the whole quantum payload is discarded. It is easy to show the factor $K$ in Eq.~\eqref{eqn:secret-key-rate} is given by,
\begin{equation}\label{eqn:parameter K}
K=P^n \dfrac{T_Q-nT_P}{T_Q},
\end{equation}
where $T_Q$ is the temporal length of the quantum payload, and $T_P$ is the processing time of the classical header at each node.

\begin{figure}[t]
    \includegraphics[]{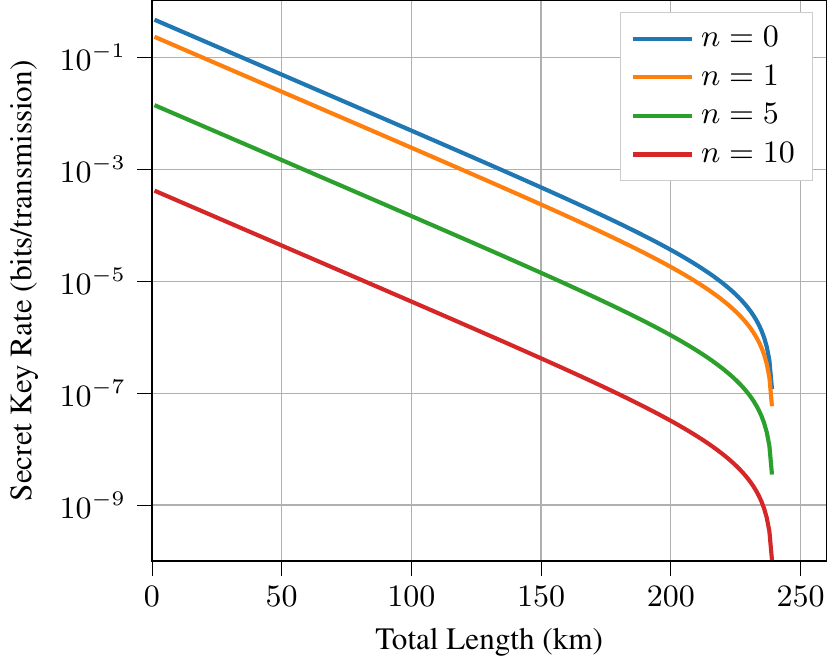}
  \caption{The secret key rates of QKD over a packet-switched quantum network without quantum memories. $n$ is the number of hops. The decrease of the secret key rate with the increase of $n$ is due to the data loss at the routers. The case of $n=0$ is equivalent to circuit switching.} 
  \label{fig:QKD}
\end{figure}

With this model, we determine the secret key rate for a varying number of mid-way switches and an independent length parameter $L$. The simulation results of the secret key rate as a function of the total channel length for different $n$ are shown in Fig.~\ref{fig:QKD}. Here we assume the quantum channels are single-mode fibers with an attenuation coefficient of
$0.2$ dB/km. Other simulation parameters are as follows: the detector efficiency is set to $0.5$; the detector dark count probability is $10^{-6}$; $f=1.15$; $T_Q=100\cdot T_P$; and $P=0.5$. The first three are typical values in QKD demonstrations. To reduce the data loss due to the processing time of classical header, we choose a relatively large ratio of  $T_Q/T_P$. The simulation results suggest that it is possible to conduct QKD over packet-switched network without using quantum memory. We remark that in a more complicated network setup, the QKD performance could be further improved by storing the quantum payload at the switch when the outgoing channel is not available. We leave this scenario as a future research for this application, and explore it further in the next section with entanglement distribution. 

\subsection{Entanglement Distribution}
Entanglement distribution is another key application enabled by quantum networks. Quantum entanglement is a resource that can be directly used in various ways to perform particular quantum tasks like quantum state teleportation \cite{bennett1993teleporting} and entanglement based quantum key distribution protocols \cite{ekert1992quantum}. Moreover, entanglement can be used as a supplemental resource within protocols to boost the performance, such as in entanglement-assisted communication~\cite{bennett1999entanglement} or various multi-party nonlocal games, one well known example being the CHSH game~\cite{clauser1969proposed}. Entanglement alone cannot be used to transmit information---known as the no-communication theorem---but used as a resource, can be used to enable quantum communication at large scale \cite{wehner2018quantum}.

An entanglement resource is composed of multiple quantum systems, each part of which sharing a correlation with the other parts. The task of distributing entanglement is therefore to transfer each communicating party a piece of the entangled system. Once transferred, they can use the entanglement as they like, to perform quantum teleportation, entanglement swapping for extending the reach of an entangled pair \cite{zukowski1993event}, an entanglement-based QKD, or any other protocol where entanglement is used. Entanglement, similarly as discussed with QKD resources, is a resource that when lost in transmission has little consequence---another entangled system can be generated and the distribution reattempted. Also, since entanglement alone contains no information by itself, and due to the monogamy of entanglement \cite{osborne2006general}, losing the resource cannot leak any secret information or correlation to potentially malicious third parties.

Along the same lines of the previous subsection, we analyze how well hybrid classical-quantum frames composed of entangled systems as payloads can be transmitted over packet-switched quantum networks. The entanglement we consider in this initial work are Einstein-Podolsky-Rosen (EPR) pairs. EPR pairs are two-qubit, maximally entangled qubits, meaning they are perfectly correlated (or anti-correlated) in measurement outcome. To quantify the entanglement-distribution ability of a packet-switched quantum network, we use the entanglement fidelity as a metric, measuring the fidelity in various network settings to observe the hop-by-hop noise effects. In simulation, the fidelity can be calculated numerically without statistical fluctuation, but naturally in an experimental setting would require more effort. Nonetheless, as a preliminary step, simulating the noise effects plays an important role for taking the first steps towards realizing a packet-switched quantum network able to distribute quantum entanglement.

Related to this section are two prior works that investigate the memory effects on the ability to distribute quantum entanglement. Instead of an approach based on optical switching, the two works \cite{kozlowski2020designing, semenenko2021entanglement} use an approach of creating pair-wise entanglement and performing entanglement swapping. In \cite{kozlowski2020designing}, Kozlowski et al. consider a network-layer protocol for managing the distribution process in a quantum network, where the hardware parameters are near-term parameters. The authors also use NetSquid, as we do, to demonstrate their protocol. Their protocol firstly establishes a virtual circuit between the sender and receiver and once a virtual circuit is established, no further routing is required. Since the protocol uses entanglement swapping as the basis for distributing entanglement, switching frames of quantum systems is not considered. In \cite{semenenko2021entanglement}, Semenenko et. al consider how varying memory lifetimes affects the entanglement generation rate in a linear chain of nodes, again, distributed via entanglement swapping, not using a packet-switching approach as we do here.

\begin{figure}
    \centering
    \includegraphics[]{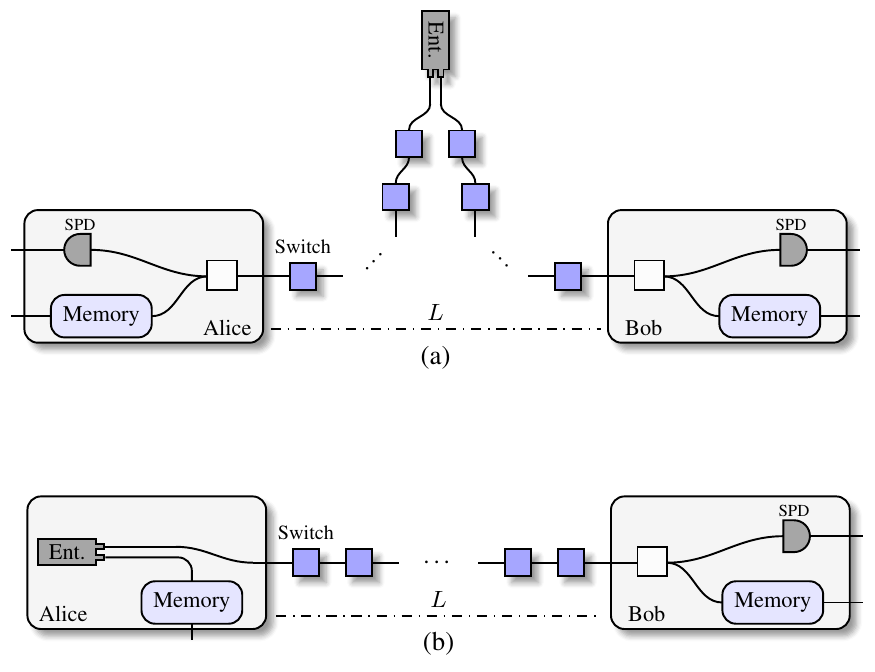}
    \caption{The network configurations used for simulating entanglement distribution.  In (a), there is a centralized node who generates pairs, sending half of an EPR in one direction, and the other half in another direction. The qubits are then routed through a linear chain of nodes. In (b), the entangled pair is generated from one of the communicating parties, storing one half of the pair in a memory, and emitting the other half through a linear chain of nodes.}
    \label{fig:ent-node-topology}
\end{figure}

\begin{figure}
    \centering
    \includegraphics[]{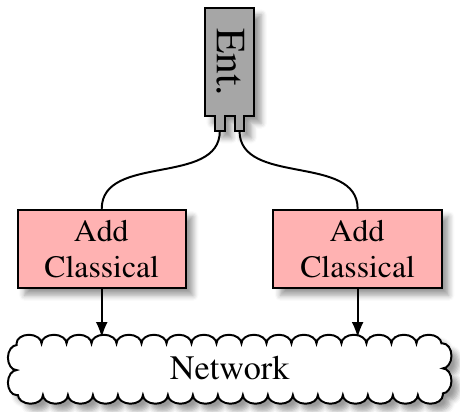}
    \caption{An entanglement source with two spacial modes emits bipartite entangled photons through two paths. Along the two paths, control information for individual receivers is added to the individual signals and then sent into the network for routing.}
    \label{fig:entanglement-distro-inner}
\end{figure}

For this simulation analysis, we determine how well the entanglement fidelity is maintained after halves of EPR pairs, a collection of which forming a frame payload, are sent through the network in various parameter regimes and network configurations. In terms of node capabilities, we use a similar configuration as in the previous section in Fig.~\ref{fig:QKD_model} except in this case, we allow the nodes in the network to have quantum memories, and the source here emits entangled pairs in two spatial modes. We consider two scenarios for entanglement distribution. In the first case, entanglement is generated at a central node and one half of each pair is sent to a respective receiver, as depicted in Fig.~\ref{fig:ent-node-topology}~(a). In this case, each path has its own hybrid frames individually prepared, and both end-nodes act as a receiver. We assume the two output paths of the quantum source act as individual quantum sources with respect to Fig.~\ref{fig:TDMWDM_new}~(a). We depict the setting in Fig.~\ref{fig:entanglement-distro-inner}, where each path gets its own header and trailer information. One can further imagine entanglement sources that output more than bipartite entanglement, extending the same concept to many output fibers. In the second case depicted in Fig.~\ref{fig:ent-node-topology}~(b), the entangled pair is generated at the sender. In this case, one half of the entangled pair is stored in a quantum memory, and the other part is sent through the network through a linear chain of nodes to a specific destination a varying number of hops away. 

The switching approach we use in this case is similar to that of the previous section, except in this case we simulate a header processing delay which delays the forwarding of any arriving qubits for the time between the header arrival to a later time when the header information is processed. Each node has a defined processing delay so that it does not forward any arriving qubits during this time. In practice, the qubits that do arrive during the processing time are stored in a memory, but those arriving after the processing time travel through to the next hop in the path without entering the memory. In this simulation, the minimum storage time is the time a full frame takes to arrive. When one half of an entangled pair reaches a receiver, the fidelity of the pair is computed. In practice, the entanglement units could be stored to perform quantum teleportation for example, which we plan to analyze more deeply in future work.

To create the simulation according to the above specifications, we use the quantum network simulation platform NetSquid \cite{coopmans2021netsquid}. NetSquid is a discrete event simulation tool that can be used for simulating complex quantum networks. The simulated nodes in the network can have various components such as quantum sources, detectors, memories, processors, and channel models such as optical fiber. Each component used in NetSquid has a physical model attached to it so to simulate noise and loss in a realistic way.

To implement the behavior of the nodes, we implement the necessary protocols in the language of NetSquid. At a high level, the three key protocols we implement are a frame-sending protocol, a frame-relaying protocol, and a frame-receiving protocol. The frame-sending protocol generates qubits from---in this case---a perfect quantum source, generating EPR pairs in the $\ket{\Psi}\coloneqq\sfrac{1}{\sqrt{2}}(\ket{00}+\ket{11})$ state. In our setting, the nodes have the ability to communicate classically and quantumly, that is, they can send both purely classical messages, for the header and trailer, and messages of quantum states which are noisy, for the payload. Before the first EPR pair is generated, a classical header message is sent across a channel. Following this, a payload worth of EPR pair halves is sent across the channel. For a predefined payload size, once the size limit of generated EPR pairs is reached, a termination trailer signal is sent signaling the end of the frame. 

The behaviour of the relay nodes is that they firstly await the classical header message. Once it arrives, they pause for the maximum between the duration of the processing time or the expected time for the total number of qubits to arrive, while simultaneously qubits are arriving in the memory. Once the processing delay elapses, the relay node firstly relays the classical header message to the next hop and begins to send the qubits in the memory onward. The qubits in the memory are emitted at the same rate they were generated in the order they arrived. At the destination node, a header message arrives followed by the qubits. The destination node in this case simply measures (in a simulated way) the fidelity of the system. Fidelity is computed using,
\begin{align}
    F\left(\rho\right) \coloneqq \left(\Tr\sqrt{\sqrt{\rho}\ketbra{\Psi}\sqrt{\rho}}\right)^{2},
\end{align}
where $\rho$ is the density matrix for a pair of entangled qubits at the time of arrival. 

The noise models used in the simulation are the following. For the depolarizing noise model of the fiber we use the following model motivated by \cite[Eq. 18]{burns1983depolarization},
\begin{align}
    p_{depol}(L, p_L) = 1 - 10^{-L \cdot p_L}. \label{eq:depol-model}
\end{align}
For the memory noise, we use the built-in \verb|T1T2NoiseModel| NetSquid model which is based on the memory properties with a $T_1$ relaxation time and a $T_2$ dephasing time. This model is explained in detail \cite[Eq. 1]{coopmans2021netsquid}. For this simulation analysis, we ignore effects of loss, and only focus on the noise. Future work will be to consider entanglement distribution rates where each the channels are lossy. To coordinate the two parties in a lossy setting would require a more complex protocol, where here we aim mainly to focus on the depolarizing effects of the channel.

\begin{figure}
    \centering
    \includegraphics[]{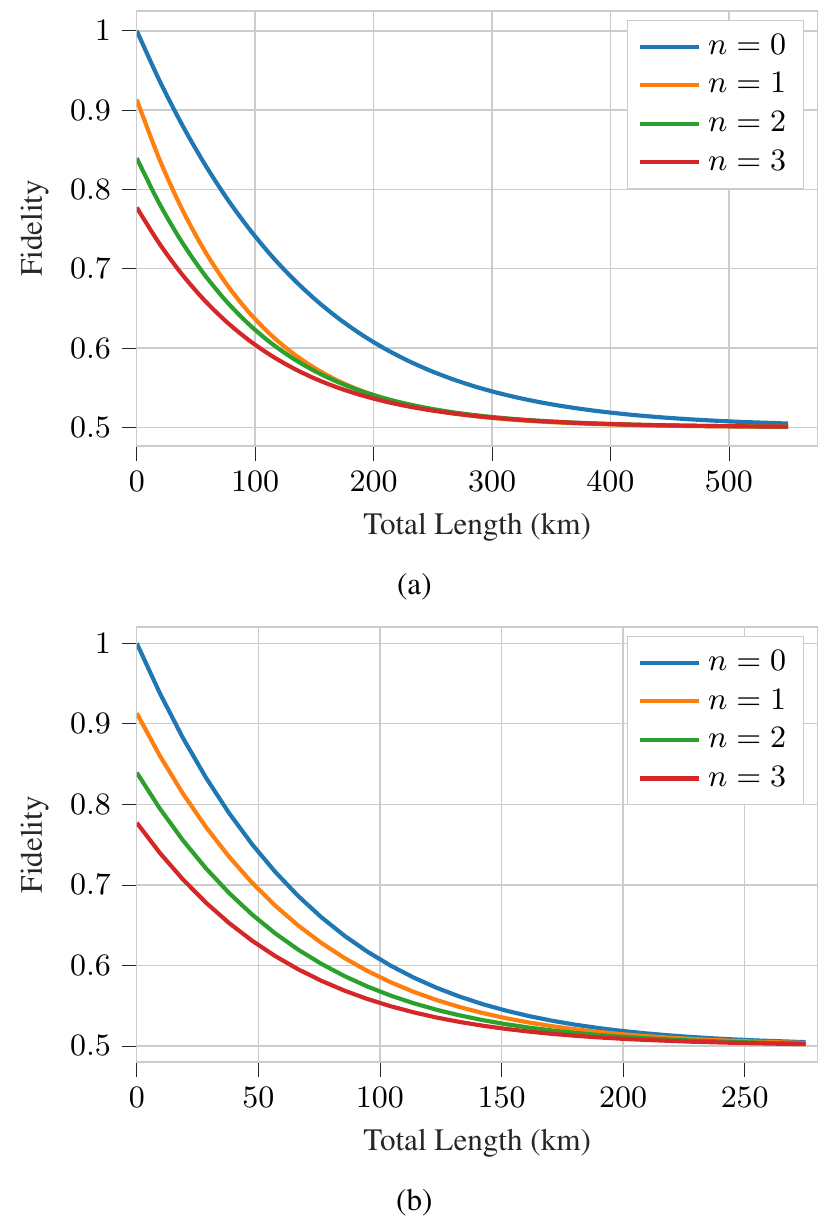}
    \caption{The effect on fidelity with respect to the number of hops in the line network. Here we fix $T_1=T_2=0.5$ ms and processing time is set to $125~\mu$s. Qubits are emitted from the source at a frequency of $5~\mu$s and $p_L=0.008$. 10 qubits per frame,  $5~\mu$s per pair. Plot (a) is for the split transmission and plot (b) is for the sender source.}
    \label{fig:fid-multi-hop}
\end{figure}

\begin{figure}
    \centering
    \includegraphics[]{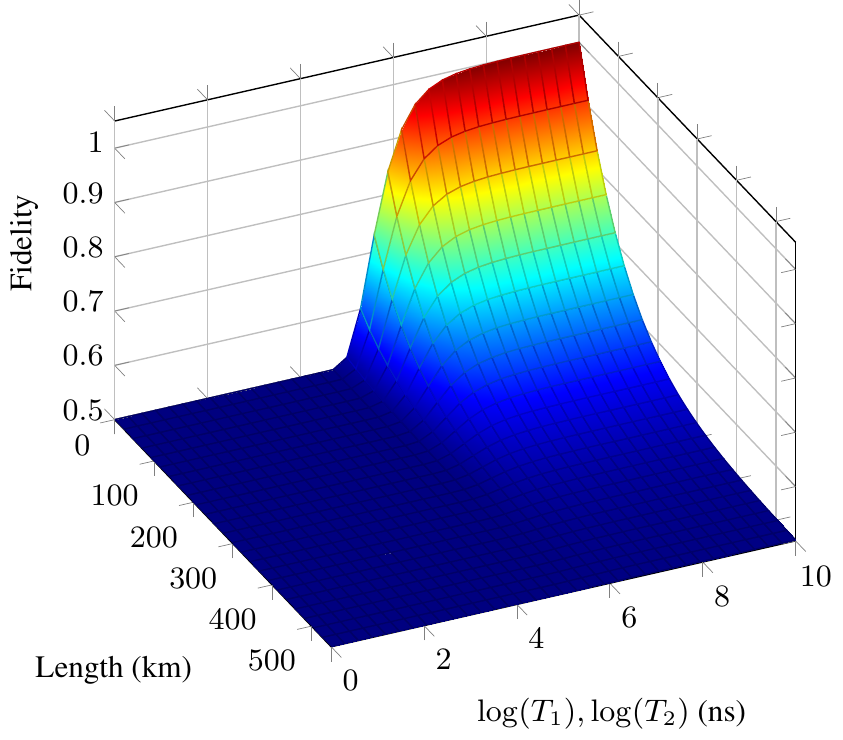}
    \caption{A comparison of fidelity effects for varying $T_1 = T_2$ and length for distributing entanglement over three hops in a network. Here, the data processing time is chosen as $125~\mu$s so that all qubits arrive in a memory. The plot is for a centralized EPR source with a split routing.}
    \label{fig:3d-ent-fid-3-hops-len}
\end{figure}

\begin{figure}
    \centering
    \includegraphics[]{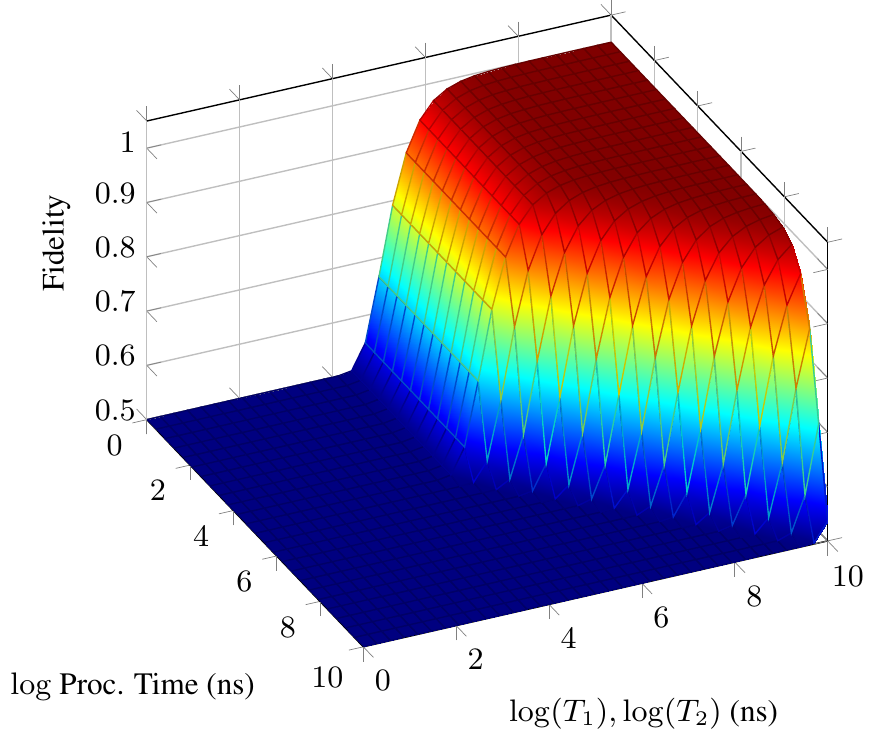}
    \caption{A comparison of fidelity effects between the header processing time and varying $T_1 = T_2$ for network 3 hops. There are 20 kms of fiber per hop, where three hops are used, and no fiber noise is used, that is $p_L=0$. The plot is for a centralized EPR source with a split routing.}
    \label{fig:3d-proc-time-t1}
\end{figure}

We begin with the analysis of the network configuration in Fig.~\ref{fig:ent-node-topology}~(a). In the upper plot of Fig.~\ref{fig:fid-multi-hop}, we see the fidelity of the entanglement with respect to the total end-to-end distance of the two nodes sharing the entanglement, where the distance ranges from $0$ km to $550$ km. The choice of parameters are $T_1=T_2=0.5$ ms, pessimistically based on results from \cite{cho2016highly}, $p_L=0.008$ defined in \eqref{eq:depol-model} to approximately match experimental distances \cite{inagaki2013entanglement}, 10 qubits are sent per frame for good simulation performance, and an EPR pair is generated every $5~\mu$s for the frame.  The header-processing delay is set to $125~\mu$s, a mid- to higher-value based on control plane stability values in burst-switched networks~\cite{barakat2007control}. In the plots of Fig.~\ref{fig:3d-ent-fid-3-hops-len} and Fig.~\ref{fig:3d-proc-time-t1}, we show how the $T_1$ and $T_2$ times affect the entanglement fidelity within a network with 3 network hops during frame transmission. In Fig.~\ref{fig:3d-ent-fid-3-hops-len}, we use the same parameters as above, except here we vary the $T_1$ and $T_2$ time and set $T_1=T_2$. In Fig.~\ref{fig:3d-proc-time-t1}, we vary the header processing time and use no depolarizing noise in the fiber (i.e. $p_L=0)$.

We interpret the results as follows. For the upper plot in Fig.~\ref{fig:fid-multi-hop}, we see, as expected, a general worsening of entanglement fidelity with the increase in total end-to-end length. As the entangled pairs enter more memories en route, indeed the fidelity of the pairs diminishes. The rate of diminishment is the smallest for a direct transmission of the pairs without any relays, where in this case, the only depolarizing noise felt by the pairs is in the fiber. At around $550$ km, in all cases the fidelity of the EPR pair falls to $0.5$, rendering it unusable. In the cases with hops, indeed the $0.5$ fidelity mark occurs at a shorter distance of roughly $400$ km. In the plot of Fig.~\ref{fig:3d-ent-fid-3-hops-len}, we see that at roughly $10^6$~ns, or $1$~ms, is when the coherence times become large enough that the fidelity begins to rapidly improve, indeed within the next 1 or 2 orders of magnitude, the fidelity already converges to a maximum for any given distance. We see, aligned with the previous results, eventually the polarizing noise of the fiber becomes too strong with the distance and the entanglement fidelity falls to 0.5 at around 500 km in the best cases. The processing time effects are seen in Fig.~\ref{fig:3d-proc-time-t1}. For small processing times in the nanosecond scale, the requirements of the memory are relatively constant. With microsecond storage times, varying of the processing time up to microsecond scale plays no role on the fidelity of the entanglement. At a larger processing times in the microsecond scale, we see that the processing time begins to require larger demand of the memories, where memory times increase directly proportionally to the increasing processing time.

\begin{figure}
    \centering
    \includegraphics[]{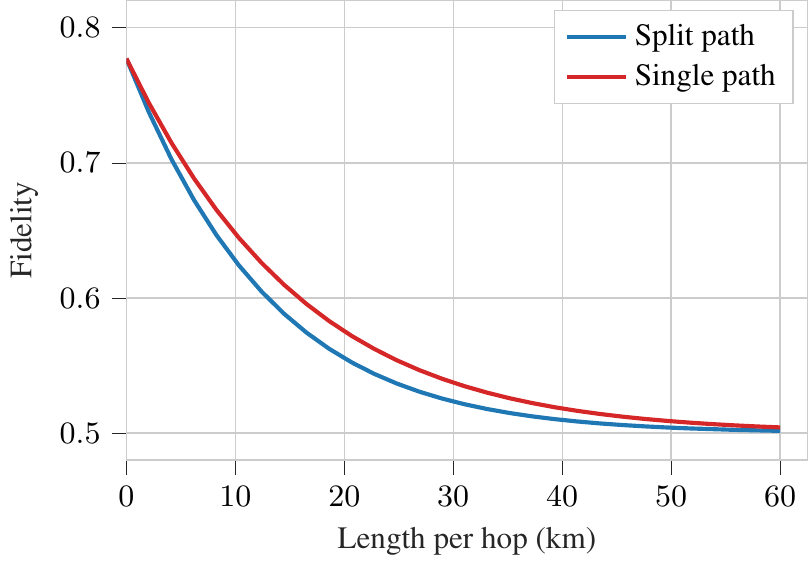}
    \caption{Comparison between the two entanglement distribution settings. Here we fix $T_1=T_2=0.5$ ms and processing time is set to $125~\mu$s. Qubits are emitted from the source at a frequency of $5~\mu$s and $p_L=0.008$. 10 qubits are sent per frame and a rate of $5~\mu$s per pair.}
    \label{fig:compare}
\end{figure}

For the second scenario in Fig.~\ref{fig:ent-node-topology}~(b), we analyse the same settings. The general trends follow the same patterns as in the previous network setting, but here, the distance in which fidelity is maintained is about half as far. In the lower plot of Fig.~\ref{fig:fid-multi-hop}, the number of hops affects the fidelity differently than in the case with a centralized source. Here only one part of the entanglement travels across the fiber, and therefore only one part experiences the depolarizing noise of the fiber. On the other hand, the other part of the entanglement is stored in a memory for the entire duration of the transmission, requiring a higher quality demand of the memory, and we see the total end-to-end distance is halved in terms of usable EPR pairs with fidelity above 0.5. The trends for the analogous plots for Fig.~\ref{fig:3d-ent-fid-3-hops-len} and Fig.~\ref{fig:3d-proc-time-t1} are excluded since for three hops, the trends vary only slightly. We plot the two fidelity trends between the two cases in Fig.~\ref{fig:compare} in the simulation setting used for Fig.~\ref{fig:fid-multi-hop}. Indeed comparing hop-by-hop distance, the noise effects are similar, and only the total lengths will vary in the plots. The single path case performs better hop-by-hop in this parameter regime, but is limited to a shorter distance since only one part of the entanglement travels.

\section{Integrating Packet-Switched Quantum Networks}\label{sec:layers}
The previous sections have been about packet switching at the link-layer level. We defined a hybrid quantum Ethernet frame structure based on LLDP frames and discussed switching approaches, analyzing some near-term applications that can be deployed in this model. In order to scale quantum networks up to the network-of-networks level, the best practices used in conventional communication networks can offer much guidance. We therefore have a good sense that for large scale general-purpose quantum inter-networks, the higher network layers need to be adapted for scalability. Still though, in the near-term, it is possible to stay at the link-layer level to support multiple users at the metropolitan scale.

Indeed, proposals for quantum network stacks motivated by the Open Systems Interconnection (OSI) model \cite{zimmermann1980osi} have been proposed \cite{pirker2019quantum, dahlberg2019link} (also see \cite{illiano2022quantum} for an in-depth comparison). These network stacks generally focus on robust entanglement distribution, which is then used to perform quantum teleportation as a means of quantum data transfer. Our design aims for a future network where quantum and classical networks can be deployed over the same network with a unified software and hardware stack. Finally, we describe a near-term approach to overcoming scalability issues regarding network utilization and communication distance in other quantum network models with an integrated packet-switched approach.

\subsection{Integration with Classical Optical Networks}

In order to integrate our packet switching of hybrid frames approach within current optical network switches, certain measures will need to be taken in order merge the two signal types. Even in our near-term vision, changes to the physical layer will need to be made to accommodate quantum signals being transmitted over the same network as classical signals. Here we describe some key issues that will need to be overcome when designing the first hybrid classical-quantum network. To overcome these issues efficiently and robustly will be the topic of future work.

Firstly, we will face the issue that quantum signals cannot be amplified. In classical ROADM implementations, optical amplification is employed at both the ingress and egress to compensate the loss of optical fiber and optical components inside the ROADM \cite{wang2019end}. Indeed removing this feature would not be an option as it would be extremely detrimental to the classical communication quality of service. Therefore methods for bypassing any optical amplifiers and classical repeaters along the fiber and inside the switches will be required. 

Next, if we aim to have quantum error correction ability as a feature of hybrid switches, the geographical distance between switches will have to be greatly reduced, or additional quantum repeater nodes will need to be introduced between switches. To avoid any detrimental effects on the classical communication signals, methods for bypassing the quantum repeaters along the fiber or disable certain quantum functions inside the q-ROADM will also be required. 

Another major issue to overcome is the cross-talk when bright classical signals and weak quantum signals coexist in the same fiber. Due to nonlinear effects in optical fibers, noise photons generated by the strong classical signals can greatly reduce the signal-to-noise ratio in the quantum channel \cite{chapuran2009optical}. Ways to overcome this issue will require a combination of control-plane scheduling and potentially using novel fiber technologies such as hollow-core fiber. In hollow-core fiber, the nonliner effects are extremely small since the signals are travelling through air rather than glass \cite{alia2022dv}.

Finally, many state-of-the-art quantum technologies can only be operated in carefully controlled environments, and it could be that switches capable of processing both classical and quantum data will need to be put in a particular container, sufficiently isolated from the environment to accommodate the quantum hardware. In this regard, some photonics-based quantum technology can already operate in room temperature \cite{krastanov2021room}, and research regarding extending the capabilities of the technologies is a large field of research and commercial interest \cite{arnold20211, moody20222022}. 

Overall, the path to completely integrated classical-quantum networks, where both the control and data plane co-exist, is a long one, and many incremental steps towards this goal will be needed. We envision that, although a full integration of classical and quantum hardware at the physical layer is unlikely, future networks could merge the hardware into a single device so to support both data types and intelligently distinguish the processing path.

\subsection{Integrating with Other Quantum Network Models}

In Section~\ref{sec:hardware}, we considered a near- and long-term versions of a quantum ROADM. In the long-term vision, the ROADM is able to correct loss- and noise-induced errors. In the short term, we consider a model only capable of storing quantum states for relatively short times. What this implies is that the near-term network model is not able to scale transmission to long distances. Indeed with technologies such as low-loss hollow-core fiber, the distance may potentially be extended \cite{poletti2014nested}, but still transmission to arbitrary distances is not possible without a form of quantum repeater. What we will investigate here is how our model of a packet-switched quantum network can be integrated with the other proposals for quantum networks. 

\begin{figure}[ht]
    \centering
    \includegraphics[]{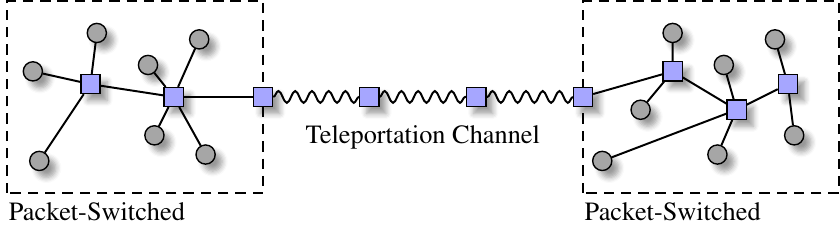}
    \caption{Two packet-switched networks are interconnected via an entanglement-based teleportation channel. Quantum data from one network is packet-switched to an egress of one subnetwork and is teleported to the ingress of another via an entanglement-based network, using entanglement swapping to establish end-to-end entanglement between the subnetwork edge nodes.}
    \label{fig:backbone-network}
\end{figure}

For quantum networks with the purpose of QKD---also called QKD networks---the integration with our model is straightforward. Current QKD networks based on fiber connections generally use trusted nodes to extend the range of key distribution. Secret keys are established point-to-point with the trusted node and relayed over encrypted channels in a hop-by-hop fashion. In this case, without quantum repeater technology, extending the range to arbitrary distances is not possible without trusted nodes. Packet switching alone cannot extend the range in this case, but we envision at metropolitan scale, with packet switching, we can increase the network utilization and throughput. In certain network traffic scenarios, low density traffic for one, packet switching offers better network performance \cite{itoh1973analysis, haas1987case}. A QKD network where the nodes regenerate keys periodically, when using AES and reducing the need to store large quantities of key material for example, is precisely this case, so we believe improvements will be seen over current circuit-switched approaches.

A packet-switched quantum network integrated with an entanglement-based network is how we envision near-term, general purpose quantum networks will inter-network. What we envision is that at a metropolitan scale, a general-purpose quantum network based on our packet-switched network model will act as a sub-network, where first-generation quantum repeater chains capable of performing entanglement swapping, would act as the backbone network, inter-connecting multiple sub-networks via a \enquote{teleportation channel}. This teleportation channel will have the sole purpose of generating long-distance entanglement to teleport quantum states over distances exceeding the reach of metropolitan scale, during the intermediate stage where quantum error correction technology is in development. With this combination of network types, we claim that scalability can be achieved in two directions---in the number of network nodes and total distance. Depicted in Fig.~\ref{fig:backbone-network} is the described structure. Moreover, based on Section \ref{sec:applications}, packet-switched based entanglement distribution can be used to implement the teleportation channel.

Using this structure, communication would work as follows: For intra-network communication---communication contained to a subnet---the packet-switched approach is used. When inter-network communication between distinct subnets is needed, packet switching is used to move the quantum information to an egress of the sub-network---a node connecting to the backbone entanglement-based network. Teleportation is used to move the information to an ingress of the destination subnet and then packet switching can again be used until the frame reaches the final destination. By merging the two network types overcoming the shortcomings of both approaches can be possible. Indeed to merge the two network types will require interoperability between the hardware and so future work will be to investigate this merger more concretely.

\section{Conclusion and Discussion}\label{sec:conclusion}

In this work, we developed a novel packet switching paradigm for quantum networks toward a quantum Internet that serves a large number of users. We introduced three properties for quantum networks: Universality, Transparency, and Scalability, on which we based our packet switching scheme. Our scheme is aligned with the paradigms of modern networks, taking a step toward a packet-switched quantum networks. We designed a classical-quantum hybrid data frame, proposed switching approaches for near- and long-term, along with initial hardware design ideas for processing and generating the frame, one of which a quantum version of a ROADM. To the data frame structure, we added additional properties in anticipation for quantum network technology capable of robustly storing and error-correcting quantum information using the already established LLDP frame structure. Next, we analyzed two quantum network use-cases suitable for near-term use in our network model, showing proof-of-concept feasibility. Lastly, we described how our model can integrate with other networks describing various predicted hurdles.

Overall, in this work we have discussed a packet-switched quantum network up to the link-layer. To increase the scale of such a network, we need to continue up the network stack into the higher layers in order to fully integrate network models towards global scale quantum networks. Moreover, the control and data plane of the network will need a deep consideration in order to integrate seamlessly with the current and coming network technologies. The state of quantum networks is much like we have experienced with classical networks, but precisely how future quantum networks will be deployed is yet to be known. With this work, we propose changing the trend of deploying quantum-only networks and instead to focus on how, with already established optical technologies, large-scale quantum networks can be integrated to classical networks, thereby unifying them.

\section*{Acknowledgements}

The authors thank Hassan Shapourian and Myungjin Lee for discussions during the development of the manuscript.


\bibliographystyle{apsrev4-2}
\input{main.bbl}

\end{document}

%% file: main.bbl
%